\documentclass[12pt,draftcls,onecolumn]{IEEEtran}
\usepackage[colorlinks=false,urlcolor=blue,citecolor=blue,linkcolor=blue,bookmarks=true, bookmarksopen=false,pdftitle=created by dvipdf,pdfcreator=NM, pdfauthor=Jain,pdfsubject=mor.sty:6.29.04]{hyperref}
\usepackage{amsmath,amsthm,amstext,amsfonts,amssymb,mathrsfs,euscript}
\usepackage{graphicx,subfigure,color,algorithm,algorithmic,fullpage,setspace}
\usepackage{dblfloatfix}
\usepackage{personal2}
\usepackage{tikz}
\usetikzlibrary{arrows,shapes}

\IEEEoverridecommandlockouts

\title{
Mechanism Designs for Stochastic Resources for Renewable Energy Integration
\thanks{The second author's research on this project is supported by the NSF CAREER award CNS-0954116 and an IBM Faculty Award.}
}

\author{
Wenyuan Tang \& Rahul Jain\\
Department of Electrical Engineering\\
University of Southern California\\
{\tt (wenyuan,rahul.jain)@usc.edu}
}
\begin{document}
\maketitle
\pagestyle{empty}

%=======================================================================================
\begin{abstract}
Among the many challenges of integrating renewable energy sources into the existing power grid, is the challenge of integrating renewable energy generators into the power systems economy. Electricity markets currently are run in a way that participating generators must supply contracted amounts. And yet, renewable energy generators such as wind power generators cannot supply contracted amounts with certainty. Thus, alternative market architectures must be considered where there are ``aggregator'' entities who participate in the electricity market by buying power from the renewable energy generators, and assuming risk of any shortfall from contracted amounts. In this paper, we propose auction mechanisms that can be used by the aggregators for procuring stochastic resources, such as wind power. The nature of stochastic resources is different from classical resources in that such a resource is only available stochastically. The distribution of the generation is private information, and the system objective is to truthfully elicit such information. We introduce a variant of the VCG mechanism for this problem. We also propose a non-VCG mechanism with a contracted-payment-plus-penalty payoff structure. We generalize the basic mechanisms in various ways. We then consider the setting where there are two classes of players to demonstrate the difficulty of auction design in such scenarios. We also consider an alternative architecture where the generators need to fulfill any shortfall from the contracted amount by buying from the spot market.
\end{abstract}

%----------------------------------------------------------------------------------------------------
\section{Introduction}

Renewable energy will increasingly constitute a greater fraction of the energy portfolio. In fact, on April 12, 2011, the Governor of California signed legislation to require one-third of the state's electricity to come from renewable energy by 2020. President ObamaÅ of 80\% electricity from clean and renewable energy by 2035 in his 2011 State of the Union address. Apart from the significant investment required, there are many system challenges in integrating renewable energy into the current power grid and electricity markets \cite{LaLo12}. These mainly arise due to the variability and unpredictability of such energy sources. For example, wind power can vary from 0 to 100 MW (at a single plant) in a matter of a couple of hours \cite{Ge09}. Solar power is equally unpredictable and highly variable, while tidal power has a cyclic nature, marked by extreme peaks (during extreme events such as storms and hurricanes).

This introduces significant challenges in matching supply with demand, which varies seasonally and by time of day but is typically quite inelastic. For this purpose, \emph{demand response} solutions are being devised where consumers are exposed to time-varying prices via smart-meters over a smart-grid infrastructure \cite{LiChLo11}. Demand response could also be done via interruptible power service contracts of varying reliability \cite{TaVa93}. Massively scalable energy storage solutions will have to be a part of the solution story if such a vision of a smart grid system is to succeed.

And yet, this is not all. With one-third of energy coming from renewable sources, challenges also arise in how the \emph{electric power economy} operates. This comprises the electricity markets that operate at various timescales (spot, day-ahead, week-ahead, etc.), the transmission capacity market, the generators, the utility companies and the consumers, and pricing to them.

Recently, a new paradigm for power system operation called \emph{Risk-Limited Dispatch} (RLD) has been proposed \cite{VaWuBi11}. The motivation is that simply building renewable energy and smart-grid infrastructure will not realize the full potential of the system. Smart operation is required as well. The existing operating paradigm of \emph{worst case dispatch} will severely underutilize the tighter feedback available between demand and supply via smart grid communication infrastructure for a more effective integration of distributed generation, renewable energy, storage and demand management. The main concern with an overwhelming mix of renewable energy in the total is the risk of not meeting operating constraints such as power balance, voltage limits, etc. \cite{MaBiBu08} The RLD operating paradigm will use real-time information about supply and demand, taking into account the stochastic nature of renewable energy sources, and determine a \emph{risk-constrained stochastic optimal dispatch} \cite{RaBiVaWu11}. This requires accurate stochastic models for renewable energy generation by the ISOs (Independent System Operators), which is usually private information of the generators.

This leads us to an important issue: How are renewable energy generators such as of wind power to participate in and integrate with the electric power economy \cite{Ab10}? \emph{How can they participate in, say day-ahead electricity markets given the uncertainty about their generation for the next day \cite{BaWeSt02,BiPoKhRaVaWu12}? What is the right market architecture that leads to efficiency?} Though the management of uncertainty seems daunting \cite{BoWaBeKeMi10}, if the market is structured the right way, enabling hundreds and thousands of geographically dispersed renewable energy generators to aggregate, then the statistical multiplexing gains can make it much easier as the variance goes down.

In this paper, we focus on the problem of aggregating power generated by renewable energy generators. It is likely that the current (day-ahead) electricity market architecture will not change, but new entities called ``aggregators'' will be allowed to enter the market, who will buy power from the renewable energy generators and then sell it in the market, assuming any risk arising due to the uncertainty in supplying the contracted amount. We investigate design of incentive mechanisms that aggregators can use in buying power from the renewable energy generators. These auctions must be designed in such a way that it induces the generators to reveal the true stochastic models/distributions of the power they will generate the next day. This then provides the right information to the aggregators to be able to plan optimally for the risk they assume due to shortfall in meeting their generation commitment in the day-ahead electricity market.

\subsubsection*{Literature Survey}
Auction and market design for electricity markets is a well-studied problem \cite{KiSt04,St02}. However, the problem we introduce in this paper which involves stochastic resources (e.g., the electricity from renewable energy sources to be supplied the next day) is new. Almost all of economic and auction theory deals with classical goods, i.e., non-stochastic goods that can be exchanged with certainty. In contrast, if a renewable energy generator contracts to supply $Q$ MW of power the next day, it may be able to supply that only with some probability $p$. With probability $1 - p$, it may fail to supply the contracted amount. The auction design problem we introduce is for such stochastic goods, which has received only scant attention, if at all.

We now provide a brief overview of some relevant work on auction and market design for electricity markets though all of it deals with classical goods. Green \cite{Gr96} studied a linear supply function market model to investigate methods of increasing competition (which leads to reduction in dead-weight losses) in the power market in England and Wales. Baldick, et al \cite{BaGrKa04} generalized Green's model by using affine functions and introducing capacity limits. Such work primarily focused on computational approaches to finding the supply function equilibrium. Johari, et al \cite{JoTs11} proposed market mechanisms based on revealing a class of supply functions parameterized by a single scalar, which is closely related to Kelly's proportional allocation mechanism. These mechanisms are designed for auctions among conventional power generators, with no uncertainty in generation. Dynamic general equilibrium models with supply friction are studied in \cite{ChMe10}.

Bitar, et al \cite{BiRaKhPoVa12} investigated how an independent renewable energy generator might bid optimally in a single competitive day-ahead electricity market with an ex-post imbalance (shortfall) penalty for scheduled contract deviations. They derived analytical expressions for the optimal contracted capacity and the corresponding expected profit. But the focus is on a decision-making problem: a single renewable generator chooses the contracted capacity to maximize his profit.
%They also studied the role of a variety of factors including improved forecasting, local generation, energy storage, etc. 

In this paper, we formulate an auction design problem for renewable energy markets. The key issue is that the generation of each renewable energy generator is a random variable. Even the player himself has no idea of the realization, but he knows the distribution. In the designed auction, we require each player to report (the parameters of) his distribution as his bid. The auctioneer then picks one or more players as the winners who have the ``best'' distributions. The main objective of auction design is to elicit the players' true types (i.e., the true distributions) so that the optimal social welfare can be achieved. We call such an auction a \emph{stochastic resource auction}. This is the first work on stochastic resource auctions to the best of our knowledge.

We propose a two-stage stochastic variant of VCG, as well as a stochastic non-VCG mechanism. We demonstrate the incentive compatibility property of the two auctions designed. Various generalizations are presented. We then consider the setting where there are two classes of players to demonstrate the difficulty of auction design in such scenarios. We also consider an alternative architecture wherein the burden of the any shortfall in generation falls on the generators themselves. We propose a VCG-type mechanism for this setting, and do its equilibrium analysis. We relegate most of the proofs to the appendix to ease readability.

%----------------------------------------------------------------------------------------------------
\section{Stochastic Resource Auction Design: Problem Statement}\label{sec:ps}

Consider $N$ renewable energy generators (player $i = 1, \ldots, N$). Player $i$'s generation for a given future time (e.g., the next day in the setting of day-ahead markets) is a random variable denoted by $X_i$, which is normalized so that $X_i \in \Om := [0,1]$. All the $X_i$'s are independent. Assume there is a type space $\Th$ such that the distribution of $X_i$ (for all $i$) can be parameterized by a $K$-dimensional vector $\th_i = (\th_i^{(1)}, \ldots, \th_i^{(K)}) \in \Th$. We refer to $\th_i$ as player $i$'s type. Let $z = (z_1, \ldots, z_N)$ be an outcome vector, where $z_i = 1$ if player $i$ is a winner and $z_i = 0$ otherwise. Denote the outcome space by $\sZ$.

\begin{definition}
A \emph{stochastic social choice function (SSCF)} $f: \Th^N \to \sZ$ for each possible profile of agents' types $\th = (\th_1, \ldots, \th_N) \in \Th^N$ specifies the outcome $f(\th) \in \sZ$.
\end{definition}

Note that the SSCF is independent of the realization $X = (X_1, \ldots, X_N) \in \Om^N$, because our goal is simply to maximize the expectation of some function of the realizations. We next define the auction form for stochastic resources as follows.

\begin{definition}
A \emph{stochastic resource auction} $M = (S, g, p, q)$ is specified by
\begin{enumerate}
\item
The strategy profile space $S = \times_{i=1}^N S_i$ from which players report their bids $s = (s_1, \ldots, s_N)$;
\item
The outcome function $g: S \to \sZ$ which determines the winners;
\item
The payment $p_i: S \to \bbR$ made by the auctioneer to player $i$ before the realization of $X$;
\item
The payment $q_i: S \times \Om^N \to \bbR$ made by player $i$ to the auctioneer after the realization of $X$.
\end{enumerate}
\end{definition}

\begin{figure}
\centering
\begin{tikzpicture}[>=latex]
\draw [->] (0,0) -- (0,5.5);
\draw [->] (4,0) -- (4,5.5);
\draw [->] (0,0) -- (4,0);
\draw [<-] (0,1) -- (4,1);
\draw [<-] (0,2) -- (4,2);
\draw [->] (0,3) -- (4,3);
\draw [->] (0,4) -- (4,4);
\draw (2,0) node[above] {Bids $\hat{\th}$};
\draw (2,1) node[above] {Outcome $g(\hat{\th})$};
\draw (2,2) node[above] {Payment $p_i(\hat{\th})$};
\draw (2,3) node[above] {Realization of $X$};
\draw (2,4) node[above] {Payment $q_i(\hat{\th}, X)$};
\draw (-0.2,2.7) node[left] {Players};
\draw (4.2,2.7) node[right] {Auctioneer};
\draw (0,5.7) node[above] {Time};
\draw (4,5.7) node[above] {Time};
\end{tikzpicture}
\caption{The stochastic resource auction.}
\label{fig1}
\end{figure}
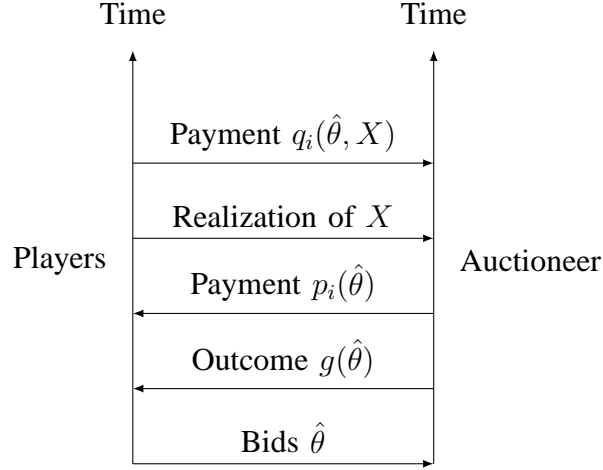

We will consider \emph{direct stochastic resource auctions} so that the strategy space $S_i$ for each player is the same as the type space $\Th$. The players report their bids $\hat{\th} = (\hat{\th}_1, \ldots, \hat{\th}_N)$ to the auctioneer. The auctioneer then determines the outcome $g(\hat{\th}) = z$, indicating the winners. A payment $p_i(\hat{\th})$ is now made to each player $i$ which only depends on the bids but not on the realization of the generation. Upon realizations, each player makes a payment $q_i(\hat{\th}, X)$ to the auctioneer. Note that a negative $p_i$ or $q_i$ indicates a reversed payment. The process of a stochastic resource auction is shown in Fig. \ref{fig1}.

Since the players are strategic, they may misreport their private information. Our goal is to design \emph{incentive compatible} mechanisms that implement the SSCF in dominant strategies, i.e., that yield truthful revelation of the players' types as the \emph{dominant strategy equilibrium}. Recall that implementation in dominant strategy equilibrium is desirable because (i) it yields greater predictability of outcome because of equilibrium uniqueness, and (ii) the informational requirements are minimal, i.e., a player does not need to know what strategies others are playing since it has a strategy that ``dominates'' (has a greater payoff) for every set of strategies played by others.

All undefined concepts are standard, and the reader can consult \cite{MaWhGr95} for their definitions. We omit their definitions and explanations due to space constraints.

%----------------------------------------------------------------------------------------------------
\section{Two Basic Mechanisms for Stochastic Resource Auctions}\label{sec:basic}

We first consider the basic scenario where there is a single winner, i.e., there exists an $i'$ such that $z_{i'} = 1$ and $z_i = 0$ for all $i \ne i'$.

Let $F_i(\cdot)$ be the cumulative distribution function (CDF) of $X_i$, which is determined by the type $\th_i$. Let $\hat{F}_i(\cdot)$ be the CDF associated with the reported type $\hat{\th}_i$.

A basic objective (represented by the SSCF) for a stochastic resource auction is to identify the player who yields the highest expected generation. We propose two designs wherein it is a dominant strategy for each player to reveal his type truthfully. The first is a variant of the well-known Vickrey-Clarke-Grove (VCG) mechanism \cite{MaWhGr95}. The second is not, but is in some sense more natural and likely to be more acceptable to generators.

%....................................................................................................
\subsection{The Stochastic VCG (SVCG) Mechanism}

We first introduce a stochastic variant of the VCG mechanism. Given the reported types, the associated distributions (and hence the expected generations) are determined. We choose the player with the highest expected generation as the \emph{winner} $i'$:
\begin{equation}\label{eq-winner}
i' \in \arg\max_i \, \int x \, \mathrm{d} \hat{F}_i(x).
\end{equation}
We also define the \emph{marginal loser} $i''$ as
\begin{equation}\label{eq-loser}
i'' \in \arg\max_{i \ne i'} \, \int x \, \mathrm{d} \hat{F}_i(x).
\end{equation}
Then, the winner makes a payment $-p_{i'}$ to the auctioneer before the realization of $X_{i'}$:
\begin{equation}
-p_{i'} = \int x \, \mathrm{d} \hat{F}_{i''}(x),
\end{equation}
which is interpreted as the contractual or sign-on amount the generator pays to the auctioneer. Upon the realization of $X_{i'}$, the auctioneer makes a payment to the winner:
\begin{equation}\label{eq-svcg-q}
-q_{i'} = X_{i'}.
\end{equation}
This can be interpreted as the payment for the supply that the generator actually makes to the aggregator (at price $1$). Then winner's payoff is
\begin{equation*}
U_{i'} = p_{i'} - q_{i'} = X_{i'} - \int x \, \mathrm{d} \hat{F}_{i''}(x).
\end{equation*}
The other players get zero payoffs. We note that the above mechanism can be easily be redefined for a general price $\mu$ (by scaling all payments by $\mu$).

\begin{theorem}\label{thm:svcg}
The SVCG mechanism specified by (\ref{eq-winner})-(\ref{eq-svcg-q}) is incentive compatible.
\end{theorem}

Due to space constraints, we omit the proof of Theorem \ref{thm:svcg}. The reader can consult the proof of Theorem \ref{thm:svcg+ssp-gen} (i) for the general case.

\noindent \textbf{Remarks.} This mechanism is a stochastic variant of the VCG mechanism. The expectation of $X_{i'}$ can be viewed as the counterpart of the valuation in the standard VCG mechanism, and $\int x \, \mathrm{d} \hat{F}_{i''}(x)$ as the counterpart of the payment (externality). Unlike the classical setting, the ``valuation'' in our model is not intrinsic. In fact, it is also in the form of a payment. This opens the possibility of other kinds of incentive compatible mechanisms.

%....................................................................................................
\subsection{The Stochastic Shortfall Penalty (SSP) Mechanism}

We now propose a non-VCG mechanism, which may be more natural to use in practice. The winner $i'$ is chosen as in (\ref{eq-winner}), and the marginal loser is defined as in (\ref{eq-loser}). A payment $p_{i'}$ is made in advance to the winner:
\begin{equation}\label{eq-ssp-p}
p_{i'} = 1.
\end{equation}
The interpretation is that the auctioneer pays for the normalized full capacity (at price $\mu = 1$) before the realization of $X_{i'}$. After realization, if there is a shortfall, the contracted generator has to pay a penalty $q_{i'}$ that depends on the shortfall:
\begin{equation}\label{eq-ssp-q}
q_{i'} =  \l (1 - X_{i'}),
\end{equation}
where
\begin{equation*}
\l = \frac{1}{1 - \int x \, \mathrm{d} \hat{F}_{i''}(x)}
\end{equation*}
can be viewed as the penalty price for the shortfall $1 - X_{i'}$. As before, the winner's payoff is
\begin{equation*}
U_{i'} = p_{i'} - q_{i'} = 1 - \l (1 - X_{i'}).
\end{equation*}
The other players get zero payoffs.

\begin{theorem}\label{thm:ssp}
The SSP mechanism specified by (\ref{eq-winner})-(\ref{eq-loser}) and (\ref{eq-ssp-p})-(\ref{eq-ssp-q}) is incentive compatible.
\end{theorem}

Due to space constraints, we omit the proof of Theorem \ref{thm:ssp}. The reader can consult the proof of Theorem \ref{thm:svcg+ssp-gen} (ii) which presents a generalization.

\noindent \textbf{Remarks.} 1. The above mechanism is inspired by \cite{BiRaKhPoVa12}, which adopts a contracted-payment-plus-penalty payoff structure for a renewable energy generator. However, they have considered a decision-making problem in which the decision maker chooses the contracted capacity, while we consider an auction design problem in which generators bid their distributions.  Note that $\l \ge 1$, which is necessary for the mechanism to be incentive compatible.
\\
2. It is interesting to observe the following ``quasi-duality'' between the SVCG and the SSP mechanisms:\\
(i) Before the realization, money flows from the winner to the auctioneer in the SVCG mechanism (which depends on the second highest bid), while it flows from the auctioneer to the winner in the SSP mechanism (which is a constant).\\
(ii) After the realization, money flows from the auctioneer to the winner in the SVCG mechanism (which depends on the realization), while it flows from the winner to the auctioneer in the SSP mechanism (which depends on both the second highest bid and the realization).

%....................................................................................................
\subsection{Revenue Comparison}

It is useful to compare the expected revenue obtained with the two mechanisms. Such a revenue comparison does make sense as the mechanisms are incentive compatible (and revenue is determined based on truthful bidding). We assume that the auctioneer resells the acquired resource $X_{i'}$ at price $\mu = 1$. The proof of the following result can be found in the Appendix.

\begin{proposition}\label{prop:rev-comp}
The auctioneer's expected revenue in the SVCG mechanism is greater than or equal to that in the SSP mechanism.
\end{proposition}

Thus, the SVCG mechanism is more favorable to an auctioneer who wants to maximize his revenue, while the SSP mechanism is more natural and useful when revenue maximization is less important.

%----------------------------------------------------------------------------------------------------
\section{Generalizations of the Basic Mechanisms}\label{sec:gen}

%....................................................................................................
\subsection{General Objective Functions}

In the basic setting, we have assumed that the social planner's objective is to contract with player $i'$ who yields the highest expected generation. Now we generalize the social planner's objective. Assume that the social planner wants to contract with the one who yields the highest $\bbE \, [h(X_i)]$, where $h: [0,1] \to \bbR$ is a function of the random variable $X_i$.

We call $h(\cdot)$ the \emph{objective function}, and we have
\begin{equation}\label{eq-obj}
\bbE \, [h(X_i)] = \int h(x) \, \mathrm{d} F_i(x).
\end{equation}
For example, the social planner's demand may be at a certain level $D \in [0,1]$. That is, he only needs $D$ amount of power and does not care about how much more would be generated. The objective function then is
\begin{equation}\label{eq-exmp1}
h(x) = \min \, \{x, D\}.
\end{equation}

A subtle point is whether the integral in (\ref{eq-obj}) converges or not. Define $\sH$ as the set of functions $h$ such that $\bbE \, [h(X)]$ exists for all $\th \in \Th$. It is natural to restrict our attention to $h \in \sH$. We also assume that the type space $\Th$ is ``well-behaved'' such that any polynomial function belongs to $\sH$; in particular, all moments of $X$ exist for all $\th \in \Th$.

Moreover, define $\sH_p := \{h \in \sH: h \ge 0, \, h(x) \le h(y), \, \forall x \le y\}$. We will show that the generalized SVCG mechanism is applicable to any $h \in \sH$, while the generalized SSP mechanism is only valid for any $h \in \sH_p$. Although $\sH_p$ is a subset of $\sH$, it still represents a large class of objective functions.

Now we propose the mechanisms. Let $i'$ denote the winner and $i''$ the marginal loser. As before, the rules of determining the winner are the same in both mechanisms:
\begin{equation}\label{eq-gen-winner}
i' \in \arg\max_i \, \int h(x) \, \mathrm{d} \hat{F}_i(x),
\end{equation}
\begin{equation}\label{eq-gen-loser}
i'' \in \arg\max_{i \ne i'} \, \int h(x) \, \mathrm{d} \hat{F}_i(x).
\end{equation}

\begin{theorem}\label{thm:svcg+ssp-gen}
(i) For any $h \in \sH$, the generalized SVCG mechanism specified by (\ref{eq-gen-winner})-(\ref{eq-gen-loser}) with
\begin{equation*}
-p_{i'} = \int h(x) \, \mathrm{d} \hat{F}_{i''}(x), \quad -q_{i'} = h(X_{i'}),
\end{equation*}
is incentive compatible.\\
(ii) For any $h \in \sH_p$, the generalized SSP mechanism specified by (\ref{eq-gen-winner})-(\ref{eq-gen-loser}) with
\begin{equation*}
p_{i'} = h(1), \quad q_{i'} = \l [h(1) - h(X_{i'})],
\end{equation*}
where
\begin{equation*}
\l = \frac{h(1)}{h(1) - \int h(x) \, \mathrm{d} \hat{F}_{i''}(x)},
\end{equation*}
is incentive compatible.
\end{theorem}

The proof of the incentive compatibility of both mechanisms is done in an integrated framework, and is available in the Appendix.

\begin{example}
It is easy to verify that the objective function in (\ref{eq-exmp1}) satisfies $h \in \sH_p \subset \sH$. Thus, both the generalized SVCG and SSP mechanisms apply. The winner's payoff in the generalized SVCG mechanism is
\begin{equation*}
U_{i'} = \min \, \{X_{i'}, D\} - \int \min \, \{x, D\} \, \mathrm{d} \hat{F}_{i''}(x),
\end{equation*}
while that in the generalized SSP mechanism is
\begin{equation*}
\begin{aligned}
U_{i'} & = \min \, \{1, D\} - \l (\min \, \{1, D\} - \min \, \{X_{i'}, D\})\\
& = D - \l (D - X_{i'})^+,
\end{aligned}
\end{equation*}
where
\begin{equation*}
\l = \frac{D}{D - \int \min \, \{x, D\} \, \mathrm{d} \hat{F}_{i''}(x)}, \quad x^+ := \max \, \{x, 0\}.
\end{equation*}
\end{example}

%....................................................................................................
\subsection{Eliminating the Undesired Equilibria}

It is worth pointing out that given an objective function $h$, reporting any $\hat{\th}_i$ whose corresponding CDF $\hat{F}_i(\cdot)$ satisfies
\begin{equation}\label{eq:eq-elim}
\int h(x) \, \mathrm{d} \hat{F}_i(x) = \int h(x) \, \mathrm{d} F_i(x)
\end{equation}
is a dominant strategy for player $i$. That is, to maximize his own payoff, player $i$ does not have to report $\th_i$ but just $\hat{\th}_i$, as long as the above equality holds.

Thus, while the proposed mechanisms are \emph{dominant strategy incentive compatible} in the sense that reporting the true type $\theta_i$ is \textit{a} dominant strategy for each player, this laxity may be considerably undesirable in some scenarios. The auctioneer may not only want to elicit the true measures for a given objective function, but also want to elicit the true types, so that this information can be used for other purposes. Hence, a stronger result would be more desirable, i.e., in addition to dominant strategy incentive compatibility, each player $i$ is strictly better off by reporting $\hat{\th}_i = \th_i$ than any $\hat{\th}_i \ne \th_i$ for some $\hat{\th}_{-i}$.

This problem can be fixed if the auctioneer does not fix (and announce) a particular objective function $h$ to be used in the mechanism but picks it arbitrarily from $\sH$ (or $\sH_p$). We thus claim the following, whose proof is in the Appendix, where further discussion is also provided.

\begin{proposition}\label{prop:eq-elim}
In the generalized SVCG (or SSP) mechanism, if the objective function $h \in \sH$ (or $\sH_p$) is chosen arbitrarily and not revealed to the bidders until the bids have been submitted, then truth-telling is a unique dominant strategy for each player.
\end{proposition}

\noindent \textbf{Remarks.} We note that the above mechanism, while \emph{incentive compatible}, is impractical, as without specifying the function $h$ to be used, the bidders may not be able to determine if their \emph{individual rationality} constraint is satisfied. Even if so, bidders may be wary of participating in an auction mechanism whose rules are not disclosed \emph{a priori}. We do note that this need not always be the case in practice as the Google AdExchange market works even though Google does not fully disclose all the rules for determining the matchings and the payments. 

%....................................................................................................
\subsection{Extension to Multiple Winners}

Now we consider the case of multiple winners. Instances of such extension of VCG mechanisms in the classical  setting can be found in \cite{Kr09}.

Suppose the social planner wants to contract with $M$ ($< N$) players who yield the highest $\bbE \, [h(X_i)]$. We rank the players in order of the bids, i.e.,
\begin{equation}\label{eq-m-winner}
\int h(x) \, \mathrm{d} \hat{F}_{i_{(1)}}(x) \ge \cdots \ge \int h(x) \, \mathrm{d} \hat{F}_{i_{(N)}}(x),
\end{equation}
where $(i_{(1)}, \ldots, i_{(N)})$ is a permutation of $\{1, \ldots, N\}$. Then players $i_{(1)}, \ldots, i_{(M)}$ are the winners and player $i_{(M+1)}$ is the marginal loser.

\begin{theorem}\label{thm:multiple}
(i) For any $h \in \sH$, the $M$-SVCG mechanism specified by (\ref{eq-m-winner}) with
\begin{equation*}
-p_{i_{(m)}} = \int h(x) \, \mathrm{d} \hat{F}_{i_{(M+1)}}(x), \quad -q_{i_{(m)}} = h(X_{i_{(m)}}), \quad m = 1, \ldots, M,
\end{equation*}
is incentive compatible.\\
(ii) For any $h \in \sH_p$, the $M$-SSP mechanism specified by (\ref{eq-m-winner}) with
\begin{equation*}
p_{i_{(m)}} = h(1), \quad q_{i_{(m)}} = \l [h(1) - h(X_{i_{(m)}})], \quad m = 1, \ldots, M,
\end{equation*}
where
\begin{equation*}
\l = \frac{h(1)}{h(1) - \int h(x) \, \mathrm{d} \hat{F}_{i_{(M+1)}}(x)},
\end{equation*}
is incentive compatible.
\end{theorem}

The proof is analogous to the previous ones and are thus omitted. The idea is to take the $(M+1)$-th highest bidder, instead of the second highest bidder, as the marginal loser; there is no difference among the $M$ winners with respect to the payment scheme.

%....................................................................................................
\subsection{Extension to Bundled Auctions}

The auctions we have considered so far only involve a single stochastic good. In smart-grid networks, an aggregator may want (or be required) to procure a mixture of different kinds of energy resources; aggregators also need to procure the (possibly stochastic) transmission capacity to transport the power. These scenarios can be modeled as the bundled stochastic resource auction, in which the social planner wants a bundle of $L$ different stochastic goods.

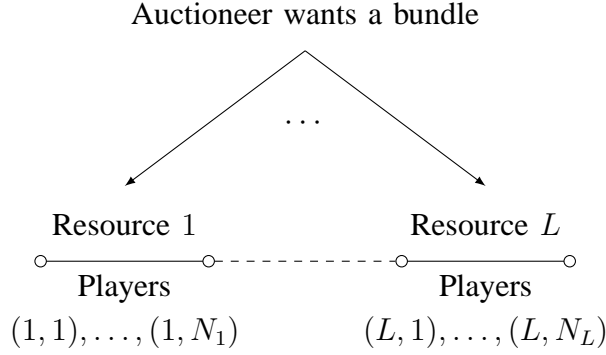
\begin{figure}
\centering
\begin{tikzpicture}[>=latex]
\draw [->] (0,0) -- (-2.4,-1.8);
\draw [->] (0,0) -- (2.4,-1.8);
\draw [o-o] (-3.6,-2.8) -- (-1.2,-2.8);
\draw [dashed] (-1.2,-2.8) -- (1.2,-2.8);
\draw [o-o] (1.2,-2.8) -- (3.6,-2.8);
\draw (0,0.2) node[above] {Auctioneer wants a bundle};
\draw (0,-1.2) node[above] {$\cdots$};
\draw (-2.4,-2.6) node[above] {Resource $1$};
\draw (-2.4,-3.5) node[above] {Players};
\draw (-2.4,-4.1) node[above] {$(1,1), \ldots, (1,N_1)$};
\draw (2.4,-2.6) node[above] {Resource $L$};
\draw (2.4,-3.5) node[above] {Players};
\draw (2.4,-4.1) node[above] {$(L,1), \ldots, (L,N_L)$};
\end{tikzpicture}
\caption{The bundled auction for stochastic resources.}
\label{fig2}
\end{figure}

To ease exposition, we consider a concrete setting of a $L$-hop network route, as shown in Fig. \ref{fig2}. The $L$ links can be viewed as $L$ goods, with $N_l$ providers for link $l$, so that there are $N := \sum_{l=1}^L N_l$ providers (players) in total. Here, link 1 could be wind energy, link 2 could be solar energy, link 3 could be hydropower, link 4 could be nuclear energy, etc. The objective in bundling them could be to ensure the right mix to satisfy certain regulatory regimes. Of course, this would also be useful for stochastic resources other than power where bundling would be more natural, e.g., network bandwidth where links might be shared with other stochastic flows.
 
We call the $i$th player on link $l$ player $(l,i)$. The resource offered by player $(l,i)$ for a future time is a random variable $X_l^i \in [0,1]$. The framework is the same as before, and player $(l,i)$'s type is $\th_l^i$ with CDF $F_l^i(\cdot)$.

We consider the case where a single winner is chosen on each link so as to maximize the expectation of a general objective function of the realizations. Let the winner on link $l$ be player $(l,i_l)$. Then the social planner's objective is to determine the set of winners that solves the following problem:
\begin{equation}\label{eq-swo}
\max_{i_1, \ldots, i_L} \, \bbE \, [h(X_1^{i_1}, \ldots, X_L^{i_L})],
\end{equation}
where the objective function $h := [0,1]^L \to \bbR$ is a function of an $L$-dimensional random vector. A useful special case is
\begin{equation}\label{eq-exmp2}
h(X_1^{i_1}, \ldots, X_L^{i_L}) = \min \, \{X_1^{i_1}, \ldots, X_L^{i_L}\},
\end{equation}
which means that the social planner wants to maximize the expected flow capacity of the entire route.

As before, define $\sH^L$ as the set of functions $h$ such that $\bbE \, [h(X_1, \ldots, X_L)]$ exists for all $(\th_1, \ldots, \th_L) \in \Th^L$. Define $\sH_p^L \subset \sH^L$ which only contains non-negative, element-wise non-decreasing functions. We will show that the bundled stochastic VCG (BSVCG) mechanism is applicable to any $h \in \sH$, while the bundled stochastic shortfall penalty (BSSP) mechanism is only valid for any $h \in \sH_p^L$. Clearly, the objective function in (\ref{eq-exmp2}) satisfies $h \in \sH_p^L \subset \sH^L$.

Moreover, we define some functions which would simplify the exposition. Assume players $(1,i_1), \ldots, (L,i_L)$ are chosen. Let
\begin{equation*}
A(i_1, \ldots, i_L) := \idotsint h(x_1, \ldots, x_L) \, \mathrm{d} \hat{F}_1^{i_1}(x_1) \cdots \mathrm{d} \hat{F}_L^{i_L}(x_L)
\end{equation*}
be the reported expected social welfare. Let
{\small
\begin{equation*}
A^l(i_1, \ldots, i_L) := \idotsint h(x_1, \ldots, x_{l-1}, X_l^{i_l}, x_{l+1}, \ldots, x_L) \, \mathrm{d} \hat{F}_1^{i_1}(x_1) \cdots\\ \mathrm{d} \hat{F}_{l-1}^{i_{l-1}}(x_{l-1}) \mathrm{d} \hat{F}_{l+1}^{i_{l+1}}(x_{l+1}) \cdots \mathrm{d} \hat{F}_L^{i_L}(x_L)
\end{equation*}
}be the reported expected social welfare, given the realization of $X_l^{i_l}$. Let
{\small
\begin{equation*}
A^{-l}(i_1, \ldots, i_L) := \idotsint h(x_1, \ldots, x_{l-1}, x_{l+1}, \ldots, x_L) \, \mathrm{d} \hat{F}_1^{i_1}(x_1) \cdots\\ \mathrm{d} \hat{F}_{l-1}^{i_{l-1}}(x_{l-1}) \mathrm{d} \hat{F}_{l+1}^{i_{l+1}}(x_{l+1}) \cdots \mathrm{d} \hat{F}_L^{i_L}(x_L)
\end{equation*}
}be the reported expected social welfare, given $X_l^{i_l} = 1$ (the normalized full capacity).

Now we propose the mechanisms for bundled auctions with general objective functions. We determine the winners $(1,i_1'), \ldots, (L,i_L')$ as follows:
\begin{eqnarray}\label{eq-bsvcg-winner}
(i_1', \ldots, i_L') \in \arg\max_{i_1, \ldots, i_L} \, A(i_1, \ldots, i_L).
\end{eqnarray}
We associate each winner $(l,i_l')$ with a set of $L$ marginal losers $(1,i_1''), \ldots, (L,i_L'')$ as follows:
\begin{eqnarray}\label{eq-bsvcg-loser}
(i_1'', \ldots, i_L'') \in \arg\max_{\substack{i_1, \ldots, i_L,\\ i_l \ne i_l'}} \, A(i_1, \ldots, i_L).
\end{eqnarray}
Therefore, winners on different links may have different sets of marginal losers (which is indeed the way our mechanisms are constructed). Also, fix a winner $(l,i_l')$ with marginal losers $(1,i_1''), \ldots, (L,i_L'')$. Clearly $i_l'' \ne i_l'$, but for any $r \ne l$, $i_r''$ is not necessarily different from $i_r'$. In particular, when the objective function $h$ can be decomposed for each link (e.g., it is a linear combination of the realizations), the winners would be chosen independently and we have $i_r'' = i_r'$ for all $r \ne l$; that is, when the winner $(l,i_l')$ on link $l$ is not present, player $(l,i_l'')$ becomes the winner on link $l$ but the winners on other links remain the same.

\begin{theorem}\label{thm:bundle}
(i) For any $h \in \sH^L$, the BSVCG mechanism specified by (\ref{eq-bsvcg-winner})-(\ref{eq-bsvcg-loser}) with
\begin{equation*}
-p_l^{i_l'} = A(i_1'', \ldots, i_L''), \quad -q_l^{i_l'} = A^l(i_1', \ldots, i_L'),
\end{equation*}
is incentive compatible.\\
(ii) For any $h \in \sH_p^L$, the BSSP mechanism specified by (\ref{eq-bsvcg-winner})-(\ref{eq-bsvcg-loser}) with
\begin{equation*}
p_l^{i_l'} = A^{-l}(i_1', \ldots, i_L'), \quad q_l^{i_l'} =  \l [A^{-l}(i_1', \ldots, i_L') - A^{l}(i_1', \ldots, i_L')],
\end{equation*}
where
\begin{equation*}
\l = \frac{A^{-l}(i_1', \ldots, i_L')}{A^{-l}(i_1', \ldots, i_L') - A(i_1'', \ldots, i_L'')},
\end{equation*}
is incentive compatible.
\end{theorem}

The proof is available in the appendix.

Note that the social welfare optimization problem in our model takes the form of (\ref{eq-swo}). If we consider another kind of system objective, say
\begin{eqnarray*}
\max_{i_1, \ldots, i_L} \, \min \, \{\bbE \, [X_1^{i_1}], \ldots, \bbE \, [X_L^{i_L}]\},
\end{eqnarray*}
then the proposed mechanisms are not applicable. However, for this special case, we can simply conduct seperate (decoupled) auctions for each link to achieve the system objective, in contrast with (\ref{eq-swo}) for which we have to conduct a single auction for the entire bundle.

Like the basic scenario, it is also straightforward to make the extension to the case of multiple winners for each good, though the notations would be quite messy. Both the VCG and the non-VCG mechanisms can be derived for this case, which we omit due to space constraints.

%----------------------------------------------------------------------------------------------------
\section{Stochastic Resource Auctions with Two Classes of Players}\label{sec:twocls}

All the models we have considered only involve one kind of players, i.e., the renewable energy generators. In reality, the generators are connected to the aggregator (auctioneer) through the transmission lines, which would incur costs for power transmission. Now we consider that the transmission lines are owned by different entities, who are also strategic players called transmission system operators (TSOs). Thus, the new scenario involves not only the generators with stochastic realizations, but also the TSOs with deterministic cost functions.

Such heterogeneity may indeed impose difficulties on the auction design. We propose a TSO-involved stochastic VCG (TSVCG) mechanism. Unlike the standard VCG mechanism, the TSVCG mechanism can only induce an efficient Nash equilibrium, a weaker equilibrium concept than a dominant strategy equlibrium.

%....................................................................................................
\subsection{Problem Statement}

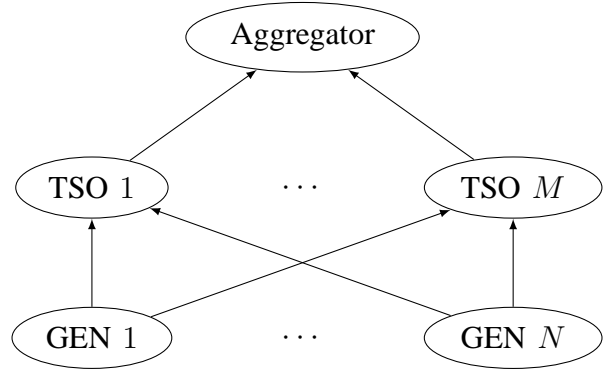
\begin{figure}
\centering
\begin{tikzpicture}[>=latex,p/.style={ellipse,draw}]
\node (A) at (0,0) [p] {Aggregator};
\node (T1) at (-2.8,-2) [p] {TSO $1$} edge [->] (A);
\node (TM) at (2.8,-2) [p] {TSO $M$} edge [->] (A);
\node (G1) at (-2.8,-4) [p] {GEN $1$} edge [->] (T1) edge [->] (TM);
\node (GN) at (2.8,-4) [p] {GEN $N$} edge [->] (T1) edge [->] (TM);
\node at (0,-2) {$\cdots$};
\node at (0,-4) {$\cdots$};
\end{tikzpicture}
\caption{The stochastic resource auction with two classes of players.}
\label{fig3}
\end{figure}

Consider $N$ generators and $M$ TSOs, which we call GEN $i$ ($i = 1, \ldots, N$) and TSO $j$ ($j = 1, \ldots, M$) respectively. There might be multiple TSOs who serve the same generator. Also, each TSO may be accessible to multiple generators (but finally can only serve one of them). That is, both $M \ge N$ and $M < N$ are possible. The topology is shown in Fig. \ref{fig3}. As before, GEN $i$'s generation is a random variable $X_i$, the distribution of which can be parameterized by a type vector $\theta_i$. Let the cost function of TSO $j$ be $c_j(x)$, where $x$ is the amount of power to be transmitted. We also assume that the cost functions can be parameterized so that the TSOs can report them exactly.

We will demonstrate that while standard VCG mechanisms might be adapted to the stochastic setting, not all the properties can be retained. To illustrate this point, we consider a simple scenario in which we pick one generator and one TSO (as a pair of winners) with the highest expected net surplus. That is, the social planner wants to contract with GEN $i'$ and TSO $j'$ such that
\begin{equation*}
(i', j') \in \arg\max_{i, j} \, \bbE \, [X_i - c_j (X_i)].
\end{equation*}

%....................................................................................................
\subsection{The TSVCG Mechanism}

Denote the reported types by $\hat{\th}_i$ for GEN $i$ and $\hat{c}_j(\cdot)$ for TSO $j$. Let GEN $i'$ and TSO $j'$ be the winners such that
\begin{eqnarray}\label{eq-tsvcg-winner}
(i', j') \in \arg\max_{i, j} \, \int [x - \hat{c}_j(x)] \, \mathrm{d} \hat{F}_i(x).
\end{eqnarray}
Let $s^{-i'}$ be the maximum expected net surplus when GEN $i'$ is not present, i.e.,
\begin{equation*}
s^{-i'} = \max_{j,i \ne i'} \, \int [x - \hat{c}_j(x)] \, \mathrm{d} \hat{F}_i(x).
\end{equation*}
Let $s^{-j'}$ be the maximum expected net surplus when TSO $j'$ is not present, i.e.,
\begin{equation*}
s^{-j'} = \max_{i, j \ne j'} \, \int [x - \hat{c}_j(x)] \, \mathrm{d} \hat{F}_i(x).
\end{equation*}
Define GEN $i'$'s payoff as
\begin{equation}
U_{i'} = X_{i'} - \hat{c}_{j'}(X_{i'}) - s^{-i'}.
\end{equation}
That is, he pays $s^{-i'}$ to the auctioneer before his actual generation is realized. Upon realization, he is paid $X_{i'} - \hat{c}_{l'}(X_{i'})$ for the supply after accounting for the cost. Define TSO $j'$'s payoff as
\begin{equation}\label{eq-tsvcg-tso}
V_{j'} = \int x \, \mathrm{d} \hat{F}_{i'}(x) - s^{-j'} - c_{j'}(X_{i'}).
\end{equation}
That is, he is paid $\int x \, \mathrm{d} \hat{F}_{i'}(x) - s^{-j'}$ by the auctioneer before the realization of $X_{i'}$. Upon realization, he incurs the intrinsic cost $c_{j'}(X_{i'})$. The other GENs and TSOs get zero payoffs.

\begin{theorem}\label{thm:tsvcg}
The TSVCG mechanism specified by equations (\ref{eq-tsvcg-winner})-(\ref{eq-tsvcg-tso}) has a Nash equilibrium in which all the GENs and TSOs report truthfully ($\hat{\th}_i = \th_i$ for all $i$ and $\hat{c}_j(\cdot) = c_j(\cdot)$ for all $j$).
\end{theorem}

The proof can be found in the appendix.

\noindent \textbf{Remarks.} It follows immediately from (\ref{eq-tsvcg-winner}) that such an equilibrium is efficient as well. However, reporting truthfully is not a dominant strategy equilibrium, since a TSO's best response would depend on the generators' strategies. Indeed, reporting truthfully is just the best response of a TSO given that the generators report truthfully, but not the dominant strategy of the TSO. This suggests the possible challenges of auction design for stochastic resources with heterogeneous players.

%----------------------------------------------------------------------------------------------------
\section{Risk-Aware Generation Assignment Auctions for Stochastic Resources}\label{sec:assign}

In previous sections, we have proposed incentive compatible mechanisms for stochastic resource auctions where the auctioneer bears the risk of not being able to acquire some capacity $Z$ from the generators. The auctioneer may be an aggregator who is committed to supplying $Z$ amount of power in an electricity market. In that case, he will have to meet any shortfall by buying power from the spot market (presumably at high prices).

Now we consider an alternative market architecture, wherein the risk of any shortfall is borne entirely by the generators themselves. That is, the generators must buy power from the spot market and deliver it to the auctioneer if there is any shortfall. This may indeed distort the incentives of the generators who may now become conservative in their bids, thus leading to inefficiency. We call such auctions \emph{risk-aware generation assignment auctions} since the generators must take the risk of the shortfall into account.

\subsection{Problem Statement}

Consider player $i$ who is required to meet a fixed demand $y_i \in [0,1]$. As before, his generation at a future time is a random variable $X_i \in [0,1]$ with CDF $F_i(\cdot)$ and probability density function (PDF) $f_i(\cdot)$. Let $\l$ be a constant denoting the price of the resource in the spot market, which can also be viewed as the ``penalty price''. Define the cost function of player $i$ as
\begin{equation*}
c_i(y_i) := \bbE \, [\l (y_i - X_i)^+] = \l \int_0^{y_i} (y_i - x) f_i(x) \, \mathrm{d} x,
\end{equation*}
which is the expected payment made by the generator to the spot market to make up for the shortfall if any, when the assigned generation is $y_i$.

We derive some properties of the cost function. It is easy to check the following:
\begin{equation*}
\frac{\mathrm{d} c_i(y_i)}{\mathrm{d} y_i} = \l \int_0^{y_i} f_i(x) \, \mathrm{d} x = \l F_i(y_i),
\end{equation*}
\begin{equation*}
\frac{\mathrm{d}^2 c_i(y_i)}{\mathrm{d} y_i^2} = \l f_i(y_i) \ge 0.
\end{equation*}
Thus, $c_i(y_i)$ is increasing and convex on $[0,1]$. Moreover,
\begin{equation*}
\left.\frac{\mathrm{d} c_i(y_i)}{\mathrm{d} y_i} \right|_{y_i = 0} = \l F_i(0) = 0,
\end{equation*}
i.e., the marginal cost at zero is zero. We also have
\begin{equation*}
\left.\frac{\mathrm{d} c_i(y_i)}{\mathrm{d} y_i} \right|_{y_i = 1} = \l F_i(1) = \l,
\end{equation*}
i.e., the marginal cost is at most $\l$, the spot market price.

Consider a stochastic demand $Z$ with PDF $f(\cdot)$ (of which a fixed demand is a special case). Likewise, we can define the aggregator's cost function (which is public information) as a function of the aggregate assigned generation $y$ ($= \sum_i y_i$):
\begin{equation*}
c(y) := \bbE \, [\l (Z - y)^+] = \l \int_y^{\infty} (z - y) f(z) \, \mathrm{d} z,
\end{equation*}
which is the expected cost of buying power from the spot market to make up for the shortfall, when the aggregate assigned generation is $y$.

It is easy to check the following:
\begin{equation*}
\frac{\mathrm{d} c(y)}{\mathrm{d} y} = -\l \int_y^{\infty} f(w) \, \mathrm{d} w = \l [F(y) - 1],
\end{equation*}
\begin{equation*}
\frac{\mathrm{d}^2 c(y)}{\mathrm{d} y^2} = \l f(y) \ge 0.
\end{equation*}
Thus, $c(y)$ is decreasing and convex on $[0,1]$. Moreover,
\begin{equation*}
\left.\frac{\mathrm{d} c(y)}{\mathrm{d} y} \right|_{y = 0} = \l [F(0) - 1] = -\l,
\end{equation*}
\begin{equation*}
\left.\frac{\mathrm{d} c(y)}{\mathrm{d} y} \right|_{y = 1} = \l [F(1) - 1] = 0.
\end{equation*}
These properties are useful in the analysis as we will see.

Our goal is to design a mechanism to make a generation assignment before the realizations of $X_i$'s and $Z$ so that the following social welfare optimization problem is solved at equilibrium:
\begin{equation}\label{swo}
\begin{aligned}
& \underset{y_1, \ldots, y_N, y}{\text{minimize}} & & \sum_i c_i(y_i) + c(y)\\
& \text{subject to} & & y = \sum_i y_i,\\
& & & 0 \le y_i \le 1, \ \forall i.
\end{aligned}
\end{equation}

Define player $i$'s payoff as $U_i(y_i) := w_i - c_i(y_i)$, where $w_i$ is the payment received by him. We consider two cases, depending on whether $f_i$'s (and hence $c_i$'s) can be parameterized or not. In the former case, each player can report the complete cost function, and we will propose a simple variant of the standard VCG mechanism. In the latter case, since it is impossible for a player to report the cost function completely, the mechanism must ask each player to communicate an approximation to the function from a finite-dimensional bid space; instead of a dominant strategy implementation (which cannot be achieved here), we seek a efficient Nash implementation in which the Nash equilibrium solves the social welfare optimization (\ref{swo}).

%....................................................................................................
\subsection{VCG Mechanism for Complete Parametrization}

In this case, each player $i$ can report the complete cost function. Denote the reported cost function by $\tilde{c}_i(\cdot)$ in the mechanism. The assignment $(y_1, \ldots, y_N, y)$ is a solution of the following optimization problem:
\begin{equation*}
\begin{aligned}
& \underset{y_1, \ldots, y_N, y}{\text{minimize}} & & \sum_i \tilde{c}_i(y_i) + c(y)\\
& \text{subject to} & & y = \sum_i y_i,\\
& & & 0 \le y_i \le 1, \ \forall i.
\end{aligned}
\end{equation*}
Let $(y_1^{-i}, \ldots, y_N^{-i}, y^{-i})$ denote the assignment as a solution of the above with $y_i = 0$, i.e., when player $i$ is not present. The payment scheme is
\begin{equation*}
w_i = \sum_{j \ne i} [\tilde{c}_j(y_j^{-i}) - \tilde{c}_j(y_j)] + c(y^{-i}) - c(y),
\end{equation*}
which is the \emph{positive externality} that player $i$ imposes on the other players by his participation.

This is just a simple variant of the standard VCG mechanism, and the incentive compatibility follows immediately.

%....................................................................................................
\subsection{i-VCG Mechanism for Incomplete Parametrization}

Now we consider the more interesting case of non-parametric $c_i(\cdot)$. In this case, it is impossible for a player to communicate the cost function exactly. Instead, we ask each player $i$ to report a two-dimensional bid $b_i = (\beta_i, d_i)$, where $\beta_i$ is his ask price and $d_i$ is the maximum quantity that he can offer at price $\beta_i$.

The assignment $(y_1, \ldots, y_N, y)$ is a solution of the following optimization problem:
\begin{equation*}
\begin{aligned}
& \underset{y_1, \ldots, y_N, y}{\text{minimize}} & & \sum_i \beta_i y_i + c(y)\\
& \text{subject to} & & y = \sum_i y_i,\\
& & & 0 \le y_i \le d_i, \ \forall i.
\end{aligned}
\end{equation*}
Let $(y_1^{-i}, \ldots, y_N^{-i}, y^{-i})$ denote the assignment as a solution of the above with $d_i = 0$, i.e., when player $i$ is not present. The payment scheme is
\begin{equation*}
w_i = \sum_{j \ne i} \beta_j (y_j^{-i} - y_j) + c(y^{-i}) - c(y),
\end{equation*}
which is the positive externality that player $i$ imposes on the other players by his participation.

We call this the i-VCG mechanism (with incomplete parametrization). Although dominant strategy implementation is impossible in this case, we show that the i-VCG mechanism is a \emph{Nash implementation}, i.e., there exists a Nash equilibrium in which the efficient assignment is achieved.

\begin{theorem}\label{thm:ivcg}
There exists an efficient Nash equilibrium in the i-VCG mechanism.
\end{theorem}

The proof can be found in the Appendix.

%----------------------------------------------------------------------------------------------------
\section{Conclusion}

In this paper, we have formulated market design problems for auctioning stochastic resources among renewable energy generators. The mechanisms can be used by an ``aggregator'' who can then bid in a futures electricity market.

We have considered two alternative market architectures. In the first, the risk due to uncertain generation is assumed by the aggregator, and the generators compete for a contract. The designed mechanisms are incentive compatible, in which the generators would truthfully reveal their probability distributions. This is achieved \emph{via} a two-part payment, an \emph{ex ante} payment (before the realization) and an \emph{ex post} payment (after the realization). Such mechanisms are important and useful for the aggregator since he can now hedge against his risk.

In the second instance, the risk due to uncertain generation is assumed by the generators, and the generators compete for assignment. If there is any shortfall, the generators are responsible for buying from the spot market and fulfilling their contracts. This can possibly skew incentives, and make the generators averse to truthfully reporting their probability distributions, as well as make achievement of social welfare optimization difficult. It turns out that this is not the case. In the parametric case, the VCG mechanism still yields incentive compatibility, while in the non-parametric case, the \emph{i}-VCG mechanism is a Nash implementation (dominant strategy implementation cannot be achieved in this case).

In the future, we will consider a repeated version of stochastic resource auctions with spot market prices that vary according to a Markov process. We would also consider a double-sided market architecture wherein there are buyers (the utility companies) as well as both types of sellers (conventional as well as renewable energy generators). It is an open question whether it is even possible to design such a market with desirable equilibrium properties. If this is impossible, then this would provide regulators with a rationale to consider alternative market architectures, as well as allow for participation of newer entities who will act as ``aggregators''.

The presented work, along with the proposed future work, will potentially provide economic solutions for integrating renewable energy generators into smart grid networks.

\section*{Acknowledgement}

The authors gratefully acknowledge suggestion of this problem by Prof. Pravin Varaiya (UC Berkeley), and the many helpful discussions they had with him, Ram Rajagopal (Stanford), Demos Teneketzis (Michigan), Mingyan Liu (Michigan) and Steven Low (Caltech).

%----------------------------------------------------------------------------------------------------

%----------------------------------------------------------------------------------------------------
\appendix

\section*{Proof of Proposition \ref{prop:rev-comp}}

\begin{proof}
Since the auctioneer gets the same amount of power in both mechanisms, his revenue from resale is the same. We just need to compare his payment to the winner. Equivalently, we can compare the payment received by the winner, which is exactly the winner's payoff. The winner's expected payoff in the SVCG mechanism is
\begin{equation*}
\begin{aligned}
\bbE \, [U_{i'}] & = \bbE \, [X_{i'}] - \int x \, \mathrm{d} F_{i''}(x)\\
& = \int x \, \mathrm{d} F_{i'}(x) - \int x \, \mathrm{d} F_{i''}(x),
\end{aligned}
\end{equation*}
while the winner's expected payoff in the SSP mechanism is
\begin{equation*}
\begin{aligned}
\bbE \, [U_{i'}] & = 1 - \l (1 - \bbE \, [X_{i'}])\\
& = 1 - \frac{1}{1 - \int x \, \mathrm{d} F_{i''}(x)} \left[1 - \int x \, \mathrm{d} F_{i'}(x)\right]\\
& = \l \left[\int x \, \mathrm{d} F_{i'}(x) - \int x \, \mathrm{d} F_{i''}(x)\right].
\end{aligned}
\end{equation*}
Since $\l \ge 1$, the winner's expected payoff in the SVCG mechanism is smaller than or equal to that in the SSP mechanism. Therefore, the auctioneer's expected revenue in the SVCG mechanism is greater than or equal to that in the SSP mechanism.
\end{proof}

\section*{Proof of Theorem \ref{thm:svcg+ssp-gen}}

\begin{proof}
We want to show that truth-telling ($\hat{\th} = \th$) is a dominant strategy equilibrium. That is, fix a player $i$ with bid $\hat{\th}_i = \th_i$, and we show that he cannot be better off by bidding $\hat{\th}_i \ne \th_i$ for any $\hat{\th}_{-i}$.

Suppose he is the winner by bidding $\hat{\th}_i = \th_i$. Let the marginal loser be $i''$. In the generalized SVCG mechanism, player $i$'s expected payoff is
\begin{equation*}
\begin{aligned}
\bbE \, [U_i] & = \bbE \, [h(X_i)] - \int h(x) \, \mathrm{d} \hat{F}_{i''}(x)\\
& = \int h(x) \, \mathrm{d} F_i(x) - \int h(x) \, \mathrm{d} \hat{F}_{i''}(x)\\
& \ge 0,
\end{aligned}
\end{equation*}
while in the generalized SSP mechanism, it is
\begin{equation*}
\begin{aligned}
\bbE \, [U_i] & = h(1) - \bbE \, [\l (h(1) - h(X_i))]\\
& = h(1) - \l \left[h(1) - \int h(x) \, \mathrm{d} F_i(x)\right]\\
& = \l \left[\int h(x) \, \mathrm{d} F_i(x) - \int h(x) \, \mathrm{d} \hat{F}_{i''}(x)\right]\\
& \ge 0.
\end{aligned}
\end{equation*}
In both mechanisms, by bidding $\hat{\th}_i \ne \th_i$, either the outcome remains the same, or he loses the contract and gets a zero payoff. Thus, he has no incentive to deviate.

Suppose he is a loser by bidding $\hat{\th}_i = \th_i$. Let the winner be $i'$. If player $i$ changes his bid so that he outbids player $i'$, his expected payoff in the generalized SVCG mechanism would be
\begin{equation*}
\bbE \, [U_i] = \int h(x) \, \mathrm{d} F_i(x) - \int h(x) \, \mathrm{d} \hat{F}_{i'}(x) \le 0,
\end{equation*}
while in the generalized SSP mechanism, it would be
\begin{equation*}
\bbE \, [U_i] = \l' \left[\int h(x) \, \mathrm{d} F_i(x) - \int h(x) \, \mathrm{d} \hat{F}_{i'}(x)\right] \le 0,
\end{equation*}
where $\l' = h(1)/(h(1) - \int h(x) \, \mathrm{d} \hat{F}_{i'}(x))$. In both mechanisms, if he does not outbid the winner, he still gets a zero payoff. Thus, he has no incentive to deviate.

This proves the theorem. Note that the condition $h(\cdot) \in \sH_p$ in the generalized SSP mechanism ensures that the penalty price $\l \ge 1$.
\end{proof}

\section*{Proof of Proposition \ref{prop:eq-elim} and Discussion}

\begin{proof}
Suppose $\hat{\th}_i$ is a dominant strategy for player $i$ for any $h \in \sH$ (or $\sH_p$). Then (\ref{eq:eq-elim}) holds for any $h \in \sH$ (or $\sH_p$). In particular, (\ref{eq:eq-elim}) holds for all $h(x) = x^n$ where $n \in \bbN$. But this exactly means that all moments of the random variables given by $\hat{\th}_i$ and $\th_i$ are the same. Therefore, $\hat{\th}_i = \th_i$. That is, truth-telling is a unique dominant strategy for each player.
\end{proof}

Actually, undesired equilibria arise due to the fact that the concept of a dominant strategy equilibrium of a mechanism is defined in a very ``weak'' sense. It requires neither that strict inequalities always holds (as in the definition of a strictly dominant strategy) nor that strict inequalities holds somewhere (as in the definition of a weakly dominant strategy).

Furthermore, the revelation principle for dominant strategies is based on such a very weak concept. In fact, analogous versions of the revelation principle (for strictly or weakly dominant strategies) do not exist. This implies that if we want to ensure that truth-telling is the unique dominant strategy, we should not restrict our attention to direct mechanisms. Proposition \ref{prop:eq-elim} provides such an alternative approach.

\section*{Proof of Theorem \ref{thm:bundle}}

\begin{proof}
We want to show that truth-telling ($\hat{\th} = \th$) is a dominant strategy equilibrium. That is, fix a player $(l,i)$ with bid $\hat{\th}_l^i = \th_l^i$, and we show that he cannot be better off by bidding $\hat{\th}_l^i \ne \th_l^i$ for any strategie profile of the other players.

Suppose he is a winner by bidding $\hat{\th}_l^i = \th_l^i$. Let the other winners be $(r,i_r')$ (for all $r \ne l$), and the marginal losers for player $(l,i)$ be $(1,i_1''), \ldots, (L,i_L'')$. We have
\begin{equation*}
A(i_1', \ldots, i_{l-1}', i, i_{l+1}', \ldots, i_L') \ge A(i_1'', \ldots, i_L'').
\end{equation*}
In the BSVCG mechanism, player $(l,i)$'s expected payoff is
\begin{eqnarray*}
\bbE \, [U_l^i] & = & \bbE \, [A^l(i_1', \ldots, i_{l-1}', i, i_{l+1}', \ldots, i_L')\\
& & - A(i_1'', \ldots, i_L'')]\\
& = & A(i_1', \ldots, i_{l-1}', i, i_{l+1}', \ldots, i_L')\\
& & - A(i_1'', \ldots, i_L'')\\
& \ge & 0,
\end{eqnarray*}
where the second equality follows from $\hat{\th}_l^i = \th_l^i$. In the BSSP mechanism, his expected payoff is
\begin{eqnarray*}
\bbE \, [U_l^i] & = & A^{-l}(i_1', \ldots, i_{l-1}', i, i_{l+1}', \ldots, i_L')\\
& & - \l \{A^{-l}(i_1', \ldots, i_{l-1}', i, i_{l+1}', \ldots, i_L')\\
& & - \bbE \, [A^l(i_1', \ldots, i_{l-1}', i, i_{l+1}', \ldots, i_L')]\}\\
& = & \l \{\bbE \, [A^l(i_1', \ldots, i_{l-1}', i, i_{l+1}', \ldots, i_L')]\\
& & - A(i_1'', \ldots, i_L'')\}\\
& = & \l [A(i_1', \ldots, i_{l-1}', i, i_{l+1}', \ldots, i_L')\\
& & - A(i_1'', \ldots, i_L'')]\\
& \ge & 0,
\end{eqnarray*}
where
\begin{equation*}
\l = \frac{A^{-l}(i_1', \ldots, i_{l-1}', i, i_{l+1}', \ldots, i_L')}{A^{-l}(i_1', \ldots, i_{l-1}', i, i_{l+1}', \ldots, i_L') - A(i_1'', \ldots, i_L'')}.
\end{equation*}
In both mechanisms, by bidding $\hat{\th}_l^i \ne \th_l^i$, either the outcome remains the same, or he loses the contract and gets a zero payoff. Thus, he has no incentive to deviate.

Suppose he is a loser by bidding $\hat{\th}_l^i = \th_l^i$. Let the winners be $(1,i_1'), \ldots, (L,i_L')$. If player $(l,i)$ changes his bid so that he outbids player $(l,i_l')$, let the new winners on the other links be $(r,i_r)$ (for all $r \ne l$). We have
\begin{equation*}
A(i_1, \ldots, i_L) \le A(i_1', \ldots, i_L'),
\end{equation*}
where $A(i_1, \ldots, i_L)$ corresponds to the strategy profile ($\hat{\th}_l^i = \th_l^i$). In the BSVCG mechanism, his expected payoff would be
\begin{equation*}
\bbE \, [U_l^i] = A(i_1, \ldots, i_L) - A(i_1', \ldots, i_L') \le 0,
\end{equation*}
while in the BSSP mechanism, his expected payoff would be
\begin{equation*}
\bbE \, [U_l^i] = \l [A(i_1, \ldots, i_L) - A(i_1', \ldots, i_L')] \le 0,
\end{equation*}
where
\begin{equation*}
\l = \frac{A^{-l}(i_1, \ldots, i_L)}{A^{-l}(i_1, \ldots, i_L) - A(i_1', \ldots, i_L')}.
\end{equation*}
In both mechanisms, if he does not outbid player $(l,i_l')$, he still gets a zero payoff. Thus, he has no incentive to deviate.

This proves the theorem. Note that the condition $h \in \sH_p^L$ in the BSSP mechanism ensures that the penalty price $\l \ge 1$.
\end{proof}

\section*{Proof of Theorem \ref{thm:tsvcg}}

\begin{proof}
The proof proceeds in two steps. First, we prove that $\hat{\th}_i = \th_i$ is a dominant strategy for each generator. Second, we prove that $\hat{c}_j(\cdot) = c_j(\cdot)$ is the best reponse of each TSO given that $\hat{\th}_i = \th_i$ for all $i$.

Fix a GEN $i$ with bid $\hat{\th}_i = \th_i$. Suppose he is a winner, and let the corresponding TSO be TSO $j$. GEN $i$'s expected payoff is
\begin{equation*}
\bbE \, [U_i] = \bbE \, [X_i - \hat{c}_j(X_i)] - s^{-i} = \int [x - \hat{c}_j(x)] \, \mathrm{d} F_i(x) - s^{-i} \ge 0.
\end{equation*}
If he changes his bid, he cannot be better off. Thus, he has no incentive to deviate. Suppose he is a loser. If he changes his bid so that he and TSO $j$ become the winners, his expected payoff would be
\begin{equation*}
\bbE \, [U_i] = \int [x - \hat{c}_j(x)] \, \mathrm{d} F_i(x) - s^{-i} \le 0,
\end{equation*}
since he is a loser when he bids $\hat{\th}_i = \th_i$. If he does not outbid the winner, he still has a zero payoff. Thus, he has no incentive to deviate.

We have shown that reporting truthfully is a dominant strategy for each GEN $i$. Given this condition, we now consider the best responses of the TSOs.

Fix a TSO $j$ with bid $\hat{c}_j(\cdot) = c_j(\cdot)$. Suppose he is a winner, and let the corresponding generator be GEN $i$. TSO $j$'s expected payoff is
\begin{equation*}
\bbE \, [V_j] = \int x \, \mathrm{d} F_i(x) - s^{-j} - \bbE \, [c_j(X_i)] = \int [x - c_j(x)] \, \mathrm{d} F_i(x) - s^{-j} \ge 0.
\end{equation*}
If he changes his bid, he cannot be better off. Thus, he has no incentive to deviate. Suppose he is a loser. If he changes his bid so that he and GEN $i$ become the winners, his expected payoff would be
\begin{equation*}
\bbE \, [V_j] = \int [x - c_j(x)] \, \mathrm{d} F_i(x) - s^{-j} \le 0,
\end{equation*}
since he is a loser when he bids $\hat{c}_j(\cdot) = c_j(\cdot)$. If he does not outbid the winner, he still has a zero payoff. Thus, he has no incentive to deviate.

This proves that there is a Nash equilibrium in which all the GENs and TSOs report truthfully.
\end{proof}

\section*{Proof of Theorem \ref{thm:ivcg}}

\begin{proof}
We have shown that $c_i(\cdot)$ is increasing and convex on $[0,1]$, and that $c(\cdot)$ is decreasing and convex. Thus, the social welfare optimization problem (\ref{swo}) is a convex optimization problem, whose solution is denoted by $(y_1^{**}, \ldots, y_N^{**}, y^{**})$. Moreover, some nice properties (such as $c_i'(0) = 0$) we have already shown ensure that the solution is interior; that is, $y_i^{**} > 0$ for all $i$, and in fact, we have
\begin{equation*}
c_1' (y_1^{**}) = \cdots = c_N' (y_N^{**}) = c'(y^{**}) =: \mu.
\end{equation*}
Consider the strategy profile: $\beta_i = \mu$ and $d_i = y_i^{**}$ for all $i$. Clearly, it induces the efficient assignment $(y_1^{**}, \ldots, y_N^{**}, y^{**})$. It remains to show that it is a Nash equilibrium.

Consider player $i$ with bid $b_i = (\mu, y_i^{**})$, whose current payoff is
\begin{equation*}
U_i (y_i^{**}) = c(y^{**} - y_i^{**}) - c(y^{**}) - c_i (y_i^{**}).
\end{equation*}
If he changes his bid to decrease his assignment by a $\delta > 0$, then the others' assignments do not change but his payoff becomes
\begin{equation*}
U_i (y_i^{**} - \delta) = c(y^{**} - y_i^{**}) - c(y^{**} - \delta) - c_i (y_i^{**} - \delta).
\end{equation*}
So his payoff changes by
\begin{eqnarray*}
& & U_i (y_i^{**} - \delta) - U_i (y_i^{**})\\
& = & [c(y^{**}) - c(y^{**} - \delta)] + [c_i (y_i^{**}) - c_i (y_i^{**} - \delta)]\\
& \le & -\mu \delta + \mu \delta\\
& = & 0.
\end{eqnarray*}
On the other hand, if he changes his bid to increase his assignment by a $\delta > 0$, then the aggregate assignment of the others decreases by $\Delta \ge 0$, and there must be $\Delta \le \delta$. His payoff becomes
\begin{equation*}
U_i (y_i^{**} + \delta) = \mu \Delta + c(y^{**} - y_i^{**}) - c(y^{**} + \delta - \Delta) - c_i (y_i^{**} + \delta).
\end{equation*}
So his payoff changes by
\begin{eqnarray*}
& & U_i (y_i^{**} + \delta) - U_i (y_i^{**})\\
& = & \mu \Delta + [c(y^{**}) - c(y^{**} + \delta - \Delta)] + [c_i (y_i^{**}) - c_i (y_i^{**} + \delta)]\\
& \le & \mu \Delta + \mu (\delta - \Delta) - \mu \delta\\
& = & 0.
\end{eqnarray*}
Thus, he has no incentive to deviate, which proves that the constructed strategy profile is a Nash equilibrium.
\end{proof}

\end{document}